\documentclass[doublecol]{epl2}
\usepackage{amsmath, amsthm, amscd, amssymb}
\usepackage{bm}
\usepackage{bbm}
\begin{document}

\title{\bf On the long time behavior of
Hilbert space diffusion}

\author{A. Bassi\inst{1,2} \and D. D\"urr\inst{2}}
\shortauthor{A. Bassi \etal}

\institute{
  \inst{1} Department of Theoretical Physics,
University of Trieste - Strada Costiera 11, 34014 Trieste, Italy\\
  \inst{2} Mathematisches Institut der L.M.U. - Theresienstr. 39, 80333
M\"unchen, Germany} \pacs{03.65.Ta}{Foundations of quantum
mechanics; measurement theory} \pacs{02.50.Ey}{Stochastic processes}
\pacs{02.30.Jr}{Partial differential equations}

\abstract{Stochastic differential equations in Hilbert space as
random nonlinear modified Schr\"odinger equations have achieved
great attention in recent years; of particular interest is the long
time behavior of their solutions. In this note we discuss the long
time behavior of the solutions of the stochastic differential
equation describing the time evolution of a free quantum particle
subject to spontaneous collapses in space. We explain why the
problem is subtle and report on a recent rigorous  result, which
asserts that any initial state converges almost surely to a Gaussian
state having a fixed spread both in position and momentum.}
\maketitle

Hilbert space valued stochastic differential equations appear all
over in quantum physics and their meaning ranges from fundamental to
effective descriptions of quantum systems. They can on the one hand
be seen as basic equations in collapse models, where the aim is to
find a unified description of microscopic quantum phenomena and
macroscopic classical ones; this is achieved by modifying the
Schr\"odinger equation, adding stochastic nonlinear terms which
model the spontaneous collapse of the wave
function~\cite{grw,col1,col1b,col2,as1,as2,as3,as4,col3,bgs}. This
fundamental meaning of Hilbert space diffusions have been originated
in a discrete version, the so called GRW model \cite{grw}, which
relies on jump processes rather than a continuous diffusion process.
One can view the diffusion process as continuum limit of the GRW
jump process~\cite{col1}; whether they can be empirically
distinguished is unclear. Mathematically, diffusion processes are
analytically easier to handle than  jump processes: that is the
reason why we consider diffusions.

The same type of equations occur in the theory of  continuous
measurement~\cite{as6,as7,meas1,meas2,meas3}; in this context, they
describe the effect of continuous measurements on the evolution of
quantum systems. Synonymous to measurement is decoherence, and
therefore the very same effective equations appear also in
decoherence theory~\cite{as5,dec}.

The  common class of stochastic differential equations used in
Quantum Mechanics has the following structure:
\begin{eqnarray} \label{eq:red-eq}
d \psi_{t} & = & \left[ -\frac{i}{\hbar}\, H\, dt + \sqrt{\lambda}
\, \sum_{n} (L_{n} - \langle L_{n} \rangle_{t}) \,
dW^{\text{\tiny $(n)$}}_{t}\right. \nonumber \\
& & - \left. \frac{\lambda}{2}\, \sum_{n} (L_{n} - \langle L_{n}
\rangle_{t})^2\, dt \right] \psi_{t},
\end{eqnarray}
with $\langle L_{n} \rangle_{t} = \langle \psi_{t}| L_{n}
\psi_{t}\rangle$. The equation is defined on a suitable domain of a
complex separable Hilbert space ${\mathcal H}$; it is manifestly
non-linear, but preserves the norm of the state vector. $H$ is the
standard quantum Hamiltonian, $\{L_{n}\}$ is a set of commuting
self-adjoint operators, $\lambda$ is a positive constant and
$\{W^{\text{\tiny $(n)$}}_{t}\}$ is a family of independent standard
Wiener process defined on a probability space $(\Omega, {\mathcal
F}, {\mathbb P})$. Eq.~\eqref{eq:red-eq} can be generalized in
several different ways, e.g. by considering non self-adjoint
collapsing operators, nevertheless the basic structure is preserved.

One of the most relevant problems in the study of this type of
equations is to analyze the  behavior of their solutions over
characteristic time regimes. The idea, which is also been put
forward in the context of the classical limit of quantum mechanics
\cite{GellMann-Hartle}, is that there are three time regimes, more
or less sharply separated, which characterize the time evolution of
the solution. 1.~\emph{Collapse regime:} That is the regime, during
which a superposition of eigenstates of $\{L_{n}\}$ collapses to one
eigenstate, with a probability close to the Born probability rule.
This feature is a consequence of the martingale
structure~\cite{col2} of the non-Schr\"odinger terms of
Eq.~\eqref{eq:red-eq}. In collapse models, the operators $\{L_{n}\}$
are functions of the position
operators~\cite{as2,as4,col1,col2,col3}, so the wave function
collapses in space; the constant $\lambda$ is chosen in such a way
that, for macroscopic systems, the collapse occurs almost
instantaneously. In decoherence models instead the collapse time is
given by the parameters characterizing the interaction with the
environment~\cite{Spohn-Duerr,Figari-Duerr-Teta}. 2.~\emph{Classical
regime:} The collapsed wave packet moves classically; the dynamical
equations are still stochastic, but the fluctuations around the
average values---which evolve classically---are very small and can
be neglected for all practical purposes~\cite{as4,col1}.
3.~\emph{Diffusive regime:} Eventually, however, the diffusive
effect of randomness takes over and the third time regime begins;
the collapsed wave packet starts to diffuse, departing from the
classical motion. In models of environment-induced collapse the
diffusion is paired with friction and  the classical trajectory goes
over into a Ornstein-Uhlenbeck type of diffusion. At the end of this
note we will come back again on these regimes, and will give
numerical estimates showing how they are characterized and separated
one from the other. The long time regime, which we are concerned
with in this paper, corresponds to the diffusive one. It is the one
which can be most easily singled out and treated rigorously.

These time regimes have been studied in particular in connection
with the following equation, defined in the Hilbert space ${\mathcal
H} = {\mathcal L}^2({\mathbb R})$:
\begin{eqnarray} \label{eq:non-lin}
d \psi_{t} & = & \left[ - \frac{i}{\hbar}\, \frac{p^2}{2m} \, dt \;
+ \; \sqrt{\lambda}\, ( q - \langle q \rangle_{t} )\, dW_{t} \right.
\nonumber \\
& & - \; \left. \frac{\lambda}{2}\, ( q - \langle q \rangle_{t}
)^2\, dt \right]\, \psi_{t};
\end{eqnarray}
$q$ is the position operator and $p$ the momentum operator. The
equation describes the time evolution of the wave function of a free
quantum particle subject to spontaneous localizations in space. Its
importance lies in the fact that it is simple enough to be analyzed
in full mathematical detail; at the same time it is physically
interesting and gives deep insight in understanding the behavior of
more complex physical situations. We mention here in particular the
results of~\cite{bs} and~\cite{bgs}: in the first case, it has been
show in mathematical detail how collapse models ensure definite
outcomes in measurements on quantum systems; in the second case, it
has been proven how the entire quantum formalism (measurable
quantities as self-adjoint operators on Hilbert spaces) follows from
the basic structure of the dynamics of collapse models. In this way
the dynamical reduction program, initiated with the original paper
of GRW, finds its completion as a consistent alternative to standard
quantum mechanics, at least at the non relativistic level.

One of the most treated problems in connection with
Eq.~\eqref{eq:non-lin}, as we said, is to establish rigorously the
long time behavior of its
solution~\cite{as2,as4,as5,as7,kol2,kol,kol3}. We report here on a
mathematical treatment of this equation where the asymptotic
behavior is fully and rigorously established~\cite{bd}, correcting
gaps in earlier works. At the end we will also discuss the three
above mentioned time regimes in physical units, and we will see how
they depend on the parameters $\lambda$ and $m$ defining the model.
The main result of~\cite{bd} is contained in the following
theorem:\vskip 0.3cm

\noindent \textsc{Theorem:} Let $\psi_{t}$ be the solution of
Eq.~\eqref{eq:non-lin}, for the given initial condition $\psi \in
{\mathcal L}^2({\mathbb R})$, $\| \psi \| = 1$. Then, with ${\mathbb
P}$--probability 1:
\begin{equation}
\lim_{t \rightarrow + \infty} \left\| \psi_{t} - \psi^{\text{\tiny
G}}_{\overline{x}_{t}, \overline{k}_{t},c_{t}} \right\| = 0,
\end{equation}
with:
\begin{eqnarray} \label{eq:gfds}
\psi^{\text{\tiny G}}_{\overline{x}_{t}, \overline{k}_{t},c_{t}}(x)
& := & \exp \left[ - \frac{z^2}{2}(x - \overline{x}_{t})^2 + i
\overline{k}_{t} x + c_{t} \right], \\
z^2 & := & (1-i)\sqrt{\frac{\lambda m}{\hbar}},
\end{eqnarray}
where $\overline{x}_{t}, \overline{k}_{t}$ are real stochastic
processes, such that
\begin{eqnarray}
\overline{x}_{t} & = & X + \frac{\hbar}{m} K\, t + \sqrt{\lambda}
\frac{\hbar}{m} \int_{0}^{t} W_{s} ds + \sqrt{\frac{\hbar}{m}} W_{t}
\nonumber \\
& & + O(e^{-\omega t/2}),
\label{eq:prima2} \\
\overline{k}_{t} & = & K + \sqrt{\lambda} W_{t} + O(e^{-\omega
t/2}), \label{eq:prima2b}
\end{eqnarray}
where $X$ and $K$ are two time-independent random variables, while
$c_{t}$ is a complex stochastic process and $\omega = 2 \sqrt{\hbar
\lambda / m}$. Of course, $c_{t}$ is such that $\psi^{\text{\tiny
G}}_{\overline{x}_{t}, \overline{k}_{t},c_{t}}$ is always correctly
normalized.\vskip 0.3cm

\noindent The above theorem states a remarkable property of
Eq.~\eqref{eq:non-lin}: with ${\mathbb P}$--probability 1 any
solution of equation~\eqref{eq:non-lin} converges in the large time
limit to a {\it Gaussian} solution having a {\it fixed spread} both
in position and momentum, while the mean in position
$\overline{x}_{t}$ as well as the mean in momentum
$\hbar\overline{k}_{t}$ undergo a random motion guided by $W_{t}$.

Instead of elaborating on the mathematical technicalities which
enter the result~\cite{bd}, we would like to explain why
mathematical rigor is needed to get to a trustable understanding.
For that, we explain first what goes wrong with ``intuitive''
arguments which have been used in the literature, which sound
convincing but which nevertheless are problematic.

An argument which has been used in~\cite{as1,as2,as3,as4,as5} to set
the asymptotic behavior of Eq.~\eqref{eq:non-lin} and of more
general equations of the type~\eqref{eq:red-eq} is based on the
large time behavior of the variance $\Delta A^2_{t} \; \equiv \; \|
(A - \langle A \rangle_{t}) \psi_{t} \|^2$ of a suitably chosen
linear operator $A$, not necessarily self-adjoint. Let us assume
that the solution $\psi_{t}$ of Eq.~\eqref{eq:red-eq} is such that
the variance, computed on this state, asymptotically vanishes:
\begin{equation} \label{eq:lim-var}
\lim_{t \rightarrow + \infty} \Delta A_{t}^2 \; = \; 0.
\end{equation}
Since $\psi$ is an eigenstate of $A$ if and only if the variance
$\Delta A^2$ is zero for such a state, it seems reasonable to conclude
that, when~\eqref{eq:lim-var} holds, $\psi_{t}$ converges to an
eigenstate of $A$, which would be the desired result. E.g., in the
case of Eq.~\eqref{eq:non-lin} the operator $A$ such
that~\eqref{eq:lim-var} is satisfied is~\cite{as4,as5}:
\begin{equation} \label{eq:gfdd}
A \; := \; q - \frac{z^2}{2\lambda m}\, p,
\end{equation}
which has precisely the coherent states $\psi^{\text{\tiny
G}}_{a,b,c}$ defined in~\eqref{eq:gfds}, with $a,b \in {\mathbb R}$,
$c \in {\mathbb C}$, as eigenstates.

The above argument is indeed true when ${\mathcal H}$ has {\it
finite} dimension and $A$ is a self-adjoint operator. Let us in fact
denote by $a_{n}$ its eigenvalues and by $\phi_{n}$ the
corresponding eigenvectors; for simplicity we assume no degeneracy
in the spectrum of $A$. From Eq.~\eqref{eq:lim-var} one trivially
has:
\begin{equation} \label{eq:lim2}
\lim_{t \rightarrow + \infty} |c_{n}(t)|^2 [ a_{n} - \langle A
\rangle_{t} ]^2 \;  = \; 0 \qquad \forall\,\, n
\end{equation}
with $c_{n}(t) = \langle \phi_{n} | \psi_{t} \rangle$. Since the
coefficients $|c_{n}(t)|^2$ sum up to 1 for any time $t$, and since
$\langle A \rangle_{t}$ has a unique limit (when it exists), then
Eq.~\eqref{eq:lim2} implies that there exists a $\overline{n}$ such
that $|c_{n}(t)|^2 \rightarrow \delta_{\overline{n},n}$ and $\langle
A \rangle_{t} \rightarrow a_{\overline{n}}$. As a consequence
$\phi_{\overline{n}}$ is the eigenstate towards which $\psi_{t}$
converges.

The above argument fails to apply in the infinite dimensional case.
As a counterexample, let us consider any orthonormal basis $\{
\phi_{n}: n \in {\mathbb N} \}$ of an infinite dimensional Hilbert
space ${\mathcal H}$, and let $A$ be a self-adjoint operator having
$\phi_{n}$ has eigenstates relative to the real eigenvalues $a_{n}$.
Let us consider the following sequence of vectors:
\begin{equation} \label{eq:ex1}
\psi_{n} \;  = \; \alpha_{n} \phi_{n} \, + \, \beta_{n} \phi_{n+1},
\qquad |\alpha_{n}|^2 + |\beta_{n}|^2 = 1.
\end{equation}
It is easy to show that $\Delta A_{n}^2 \; = \;  |\alpha_{n}|^2
|\beta_{n}|^2 \, (a_{n+1} - a_{n} )^2$, which asymptotically goes to
0 as soon as $(a_{n+1} - a_{n} ) \rightarrow 0$ for $t \rightarrow +
\infty$, without requiring that either $\alpha_{n}$ or $\beta_{n}$
asymptotically vanishes. In other words, the variance of $A$ goes to
zero while the  $\psi_{n}$ need not approach any of the eigenstates.

The  reason why the above proof breaks down in infinite dimensional
spaces is that the spectrum of an operator may have an accumulation
point, or may be continuous---which is actually the case with the
operator $A$ defined in~\eqref{eq:gfdd}---in which case the proof
for the finite-dimensional case does not hold anymore. The physical
reason instead is that, for such a kind of an operator, the
variance, though vanishingly small, is always bigger than the
difference between two eigenvalues, and thus it cannot discriminate
between the corresponding two eigenstates.

There is a second argument which has been used to study the solution
of Eq.~\eqref{eq:non-lin}, in particular its long time behavior,
which relies on its expansion in terms of coherent states. It is
known~\cite{as4,meas1,kol2,kol,kol3,jp1} that Gaussian wave
functions are solutions of Eq.~\eqref{eq:non-lin}. To see this, let
us consider the following linear stochastic differential equation:
\begin{equation} \label{eq:le}
d\,\phi_{t}(x) \; = \; \left[ -\frac{i}{\hbar}\, \frac{p^2}{2m}\, dt
+ \sqrt{\lambda}\, q \, d\xi_{t} - \frac{\lambda}{2}\, q^2 dt
\right] \phi_{t}(x),
\end{equation}
where $\xi_{t}$ is a standard Wiener process defined on a
probability space $(\Omega, {\mathcal F}, {\mathbb Q})$. The square
norm $\| \phi_{t} \|^2$ is a martingale~\cite{jp1,hol} solving the
stochastic differential equation $d \| \phi_{t} \|^2 = 2
\sqrt{\lambda} \langle q \rangle_{t}\| \phi_{t} \|^2 d\xi_{t}$, and
it can be used as a Radon-Nikodym derivative of a new probability
measure. Let us choose ${\mathbb Q}$ in such a way that $d {\mathbb
P}/ d {\mathbb Q} = \| \phi_{t} \|^2$; Girsonov's theorem implies
that the two Wiener processes $W_{t}$ and $\xi_{t}$ are related as
follows~\cite{jp1,hol}: $dW_{t} = d\xi_{t} - 2 \sqrt{\lambda}
\langle q \rangle_{t} dt$. Given these results, a straightforward
application of It\^o calculus shows that if $\phi_{t}$ solves
Eq.~\eqref{eq:le}, than $\psi_{t} = \phi_{t}/\| \phi_{t} \|$ solves
Eq.~\eqref{eq:non-lin}. In other words, one can work with
Eq.~\eqref{eq:le}---which is {\it linear}---and, at the end,
normalize the wave function and replace $\xi_{t}$ with $W_{t}$
according to the previous formula in order to get the corresponding
solution of Eq.~\eqref{eq:non-lin}, for any suitable initial state.

Direct substitution shows that Gaussian wave functions
\begin{equation} \label{eq:dfgf}
\phi_{t}(x) \; = \; \exp\left[ - \alpha_{t} (x - x^{\text{\tiny
m}}_{t})^2 + i k^{\text{\tiny m}}_{t}x + \gamma_{t} \right],
\end{equation}
solve Eq.~\eqref{eq:le}, if $x^{\text{\tiny m}}_{t}$,
$k^{\text{\tiny m}}_{t}$, $\gamma_{t}$ solve suitable stochastic
differential equations which we do not report here (see
e.g.~\cite{as4,jp1} and references therein), while $\alpha_{t}$
solves the deterministic equation:
\begin{equation} \label{eq:gxdxdfda}
\frac{d}{dt}\, \alpha_{t} \; = \; \lambda - \frac{2i\hbar}{m}
\alpha^2_{t}.
\end{equation}
This equation has the nice feature that $\alpha_{t} \rightarrow
z^2/2$ for $t \rightarrow + \infty$, for any initial condition
$\alpha_{0}$. In other words, {\it any} initial Gaussian wave
function converges, in the long time limit, to a wave function of
the type~\eqref{eq:gfds}. Since ${\mathcal L}^2$--functions can be
expressed in terms of coherent states as follows:
\begin{equation} \label{eq:gfzp}
\phi(x) = \int dx^{\text{\tiny m}}_{0} dk^{\text{\tiny m}}_{0}
f(x^{\text{\tiny m}}_{0},k^{\text{\tiny m}}_{0}) \exp\left[ -
\alpha_{0} (x - x^{\text{\tiny m}}_{0})^2 + i k^{\text{\tiny
m}}_{0}x  \right],
\end{equation}
then using linearity of Eq.~\eqref{eq:le} and the asymptotic
behavior of $\alpha_{t}$, one is tempted to conclude that also the
general solution of Eq.~\eqref{eq:non-lin} converges to a Gaussian
state of the type~\eqref{eq:gfds}. This argument has been used e.g.
in~\cite{as6,kol}. This conclusion however is false in general.

That it is false may be seen by scrutinizing the
physically interesting example of the non-linear but norm-preserving
differential equation in ${\mathcal L}^{2}({\mathbb R})$:
\begin{equation} \label{eq:loc-eq}
\frac{d}{dt} \, \psi_{t}(x) = \left[ - \frac{i}{\hbar}
\frac{p^2}{2m} - \lambda (q^2 - \langle q^2 \rangle_{t}) \right]
\psi_{t}(x).
\end{equation}
The Schr\"odinger term spreads out the wave function along the real
axis while the other two terms localize it in space (the effect of
$\langle q^2 \rangle_{t}$ is to keep the norm of $\psi_{t}$ constant
in time); these two effect are opposite to each other and thus
cancel when an equilibrium is reached. It is then reasonable to
expect that any initial wave function reaches asymptotically such an
equilibrium, i.e. it approaches a wave function whose spread is the
result of the compromise between the two competing effects. Note
that this is precisely the kind of intuition concerning the large
time behavior of the solutions of Eq.~\eqref{eq:non-lin}.

It is easy to show that a Gaussian wave functions of the
type~\eqref{eq:dfgf} are solutions of Eq.~\eqref{eq:loc-eq} if
$\alpha_{t}$ solves Eq.~\eqref{eq:gxdxdfda}, while $x^{\text{\tiny
m}}_{t}$, $k^{\text{\tiny m}}_{t}$ and $\gamma_{t}$ satisfy the
following equations:
\begin{eqnarray}
\frac{d}{dt}\, x^{\text{\tiny m}}_{t} & = & \frac{\hbar}{m}\,
k^{\text{\tiny m}}_{t} -
\frac{\lambda}{\alpha^{\makebox{\tiny R}}_{t}}\, x^{\text{\tiny m}}_{t}, \label{eq:xst}\\
\frac{d}{dt}\, k^{\text{\tiny m}}_{t} & = & 2\lambda \,
\frac{\alpha^{\makebox{\tiny I}}_{t}}{\alpha^{\makebox{\tiny
R}}_{t}} \, x^{\text{\tiny m}}_{t},  \label{eq:xst2}
\end{eqnarray}
the symbols $\alpha^{\text{\tiny R}}_{t}$ and $\alpha^{\text{\tiny
I}}_{t}$ denote respectively the real and imaginary parts of
$\alpha_{t}$. The two equations for $x^{\text{\tiny m}}_{t}$ and
$k^{\text{\tiny m}}_{t}$ imply that both these quantities vanish for
large times. We then have the following result: {\it any} initial
Gaussian solution asymptotically converges towards a Gaussian wave
function centered in the origin of both the position and the
momentum space, and with a fixed spread equal to $z^2/2$. We are
then precisely in the situation described before, which seems to
indicate that any solution of Eq.~\eqref{eq:loc-eq}---at least those
which can be initially written as in~\eqref{eq:gfzp}---converges
asymptotically to a Gaussian state with a fixed spread. However, we
show that this is not true by explicitly finding the general
solution of Eq.~\eqref{eq:loc-eq}.

Such an equation can be easily solved by first considering the
following linear equation:
\begin{equation} \label{eq:lin-loc-eq}
i \hbar\, \frac{d}{dt} \, \phi_{t}(x) = \left[ \frac{p^2}{2m} -
i\hbar\lambda\, q^2 \right] \phi_{t}(x);
\end{equation}
it is immediate to show that if $\phi_{t}$ solves
Eq.~\eqref{eq:lin-loc-eq}, then $\psi_{t} = \phi_{t}/ \| \phi_{t}
\|$ (assuming that $\phi_{t} \neq 0$) solves Eq.~\eqref{eq:loc-eq}.
The above equation describes the so called non-self-adjoint (NSA)
harmonic oscillator and has been already considered in the
literature~\cite{dav}, and its general solution is known. The {\it
eigenvalues} of the operator $H = p^2/2m - i\hbar\lambda q^2$ are:
\begin{equation} \label{eq:dssd}
\lambda_{n} = \frac{1-i}{2}\, \hbar \omega_{n}, \quad \omega_{n} =
\left( n + \frac{1}{2} \right) \omega, \quad \omega = 2
\sqrt{\frac{\hbar \lambda}{m}}
\end{equation}
and the corresponding {\it eigenvectors} are:
\begin{equation} \label{eq:ev}
u_{n}(x) \; \equiv \; \sqrt{z} \exp\left[ -\frac{z^2}{2}\, x^2
\right] \overline{H}_{n}(zx),
\end{equation}
where $\overline{H}_{n}(x)$ is the normalized Hermite polynomial of
degree $n$. The general solution of Eq.~\eqref{eq:lin-loc-eq} can
then be written as a superposition of the states:
\begin{equation} \label{eq:jvcxz}
\phi^{\text{\tiny $(n)$}}_{t}(x) \; = \; \exp\left[- \frac{1+i}{2}
\omega_{n} t\right]\, u_{n}(x),
\end{equation}
and the corresponding solution of Eq.~\eqref{eq:loc-eq} can be
obtained by dividing $\phi_{t}(x)$ by its norm; in particular, the
normalized states $\psi^{\text{\tiny $(n)$}}_{t}(x) =
\phi^{\text{\tiny $(n)$}}_{t}(x) / \| \phi^{\text{\tiny $(n)$}}_{t}
\|$ are {\it stationary} solutions of Eq.~\eqref{eq:loc-eq} for any
value of $n$, as the modulus $|\phi^{\text{\tiny $(n)$}}_{t}(x)|$ is
constant in time: this in particular means that they never approach
a Gaussian state, with the only exception of the ground state
$\psi^{\text{\tiny $(0)$}}_{t}(x)$ which is already a Gaussian
function.

One can easily understand that if $\phi_{t}$ is a superposition of
eigenstates $\phi^{\text{\tiny $(n)$}}_{t}$ with $n \in {\mathcal N}
\subseteq {\mathbb N}$, then the normalized state $\psi_{t}$
approaches for large times the stationary state
$\psi_{t}^{\text{\tiny $(\overline{n})$}}$ such that $\overline{n} =
\min \{n, n \in {\mathcal N} \}$. In fact, the non-normalized
eigenstate $\phi_{t}^{\text{\tiny $(\overline{n})$}}$ is the one
with the weakest time dependent damping factor; accordingly, when
one normalizes $\phi_{t}$ to obtain $\psi_{t}$, this is the only
term of the superposition for which the damping factor cancels out,
while all other terms keep an exponential factor proportional to
$\exp[- (\omega_{n} - \omega_{\overline{n}})t/2]$. This means that
Eq.~\eqref{eq:loc-eq} has not only one stationary state, but
infinitely many, since for any single-eigenstate solution the
damping factor is canceled by the normalization.

The above analysis disproves the physical intuition about the
equation and the argument according to which properties of Gaussian
states are sufficient to analyze the behavior of the general
solution. The reason why, in this case, Gaussian wave functions do
not provide a complete description is that they always have a
non-zero component with respect to the ground state
$\psi^{\text{\tiny $(0)$}}_{t}$ so they necessarily converge towards
it. But this excludes the situation in which the initial wave
function does have a null component on such a state. Note also that
the states $\psi^{\text{\tiny $(n)$}}_{t}$ have a bigger spread, the
bigger the value of $n$; accordingly, it is not even true that
Eq.~\eqref{eq:loc-eq} localizes wave functions in space, below a
fixed spread.

{}From a more mathematical point of view, the reason why the
analysis of the long time behavior in terms of coherent states leads
to the wrong conclusion is rather subtle; we exemplify this by
considering the first eigenstate $u_{1}(x)$. Such a state can be
expressed as in~\eqref{eq:gfzp}: the Gaussian term $\exp\left[ -
\alpha_{0} (x - x^{\text{\tiny m}}_{0})^2 + i k^{\text{\tiny
m}}_{0}x  \right]$ evolves to a state of the form~\eqref{eq:dfgf},
where $\alpha_{t}$, $x^{\text{\tiny m}}_{t}$ and $k^{\text{\tiny
m}}_{0}$ solve Eqs.~\eqref{eq:gxdxdfda},~\eqref{eq:xst}
and~\eqref{eq:xst2}, respectively, while $\gamma_{t}$ solves the
following equation:
\begin{equation}
\frac{d}{dt}\gamma_{t}\; = \;  \lambda \left[ 1 - 2\,
\frac{\alpha_{t}}{\alpha^{\text{\tiny R}}_{t}}
\right](x^{\text{\tiny m}}_{t})^2 - \frac{i\hbar}{2m}(k^{\text{\tiny
m}}_{t})^2 - \frac{i \hbar}{m} \alpha_{t}. \label{eq:jfg}
\end{equation}
The function $f(x^{\text{\tiny m}}_{0},k^{\text{\tiny m}}_{0})$ is
highly undetermined, as the Gaussian states form an over-complete
set; nevertheless it has the symmetry property: $f(x^{\text{\tiny
m}}_{0},k^{\text{\tiny m}}_{0}) = -f(-x^{\text{\tiny
m}}_{0},-k^{\text{\tiny m}}_{0})$, since $u_{1}(x) = - u_{1}(-x)$.
Then the state at time $t$ reads:
\begin{eqnarray}
\phi^{\text{\tiny $(1)$}}_{t}(x) & = & \int dx^{\text{\tiny m}}_{0}
dk^{\text{\tiny m}}_{0}\, f(x^{\text{\tiny m}}_{0},k^{\text{\tiny
m}}_{0})\, \exp \left[\gamma_{t}\right]\qquad \nonumber \\
& & \quad\quad\quad\cdot \exp\left[ - \alpha_{t} (x - x^{\text{\tiny
m}}_{t})^2 + i k^{\text{\tiny m}}_{t}x  \right];
\end{eqnarray}
since $x^{\text{\tiny m}}_{t}$ and $k^{\text{\tiny m}}_{t}$ tend to
0 asymptotically, it seems reasonable that the second exponential
can be easily taken out of the integral, so to speak. What is left
however is the integral of an odd function---as one can see from
Eq.~\eqref{eq:jfg}, the global factor $\gamma_{t}$ is even in the
initial conditions $x^{\text{\tiny m}}_{0}, k^{\text{\tiny
m}}_{0}$---which of course is zero. Therefore the large time
behavior requires much more scrutiny and cannot be read off from the
``zeroth-order approximation''.

The above discussion is useful not only because it shows that care
must be taken when analyzing the large time behavior of the
solutions of Eq.~\eqref{eq:non-lin}, but also because it helps in
understanding what actually happens in this case. As proven
in~\cite{bd} each solution solution of Eq.~\eqref{eq:le}, from which
the corresponding solution of Eq.~\eqref{eq:non-lin} can be directly
obtained, can be mapped into a solution of Eq.~\eqref{eq:lin-loc-eq}
and can be written as follows:
\begin{equation} \label{eq:fgbxcv}
\phi_{t}(x) \; = \; e^{i \overline{k}_{t} x + \gamma_{t}}
\sum_{n=0}^{+\infty} \overline{\alpha}^{\makebox{\tiny $(n)$}}_{t}
\phi^{\text{\tiny $(n)$}}_{t}(x - \overline{x}_{t}),
\end{equation}
where $\phi^{\text{\tiny $(n)$}}_{t}(x)$ are exactly the eigenstates
of the NSA harmonic oscillator defined in~\eqref{eq:jvcxz}, while
$\overline{x}_{t}$ and $\overline{k}_{t}$ are the stochastic
processes introduced in~\eqref{eq:prima2} and~\eqref{eq:prima2b},
and $\gamma_{t}$ is a new stochastic process function of $c_{t}$
appearing in~\eqref{eq:gfds}; moreover
\begin{equation} \label{eq:gfda}
\overline{\alpha}^{\makebox{\tiny $(n)$}}_{t} \; = \;
\sum_{k=0}^{+\infty} \alpha_{k+n}
\frac{\sqrt{(k+n)!}}{\sqrt{n!}k!}\, (\sqrt{2}z
\overline{\zeta}_{t})^k,
\end{equation}
$\overline{\zeta}_{t}$ being another stochastic process; the complex
terms $\alpha_{m}$ are the coefficients of the superposition of the
initial state $\phi_{0}$ in terms of the eigenstates
$\phi^{\text{\tiny $(m)$}}_{0} \equiv u_{m}$.

A crucial point is that the series defined in~\eqref{eq:fgbxcv} is
norm convergent {\it only} for $t > (4c +1)/\omega$, where $c$ is a
constant. This means that the representation of the general solution
of Eq.~\eqref{eq:le} given in~\eqref{eq:fgbxcv} is suitable only for
studying the large time behavior of $\psi_{t}$ solution of
Eq.~\eqref{eq:non-lin}, not its properties at finite small (compared
to $\omega$) times, which are also very interesting---even more
interesting than the long time behavior, from a physical point of
view---as discussed in~\cite{bd}.

As we can see from~\eqref{eq:fgbxcv}, the collapse mechanism is
basically the same as in the previous example of
Eq.~\eqref{eq:loc-eq}: $\psi_{t} = \phi_{t}/ \| \phi_{t} \|$
converges to the lowest non-vanishing eigenstate appearing in the
sum~\eqref{eq:fgbxcv}, which is the only one whose exponential
damping factor is fully canceled by normalization. However, a
crucial difference with respect to the solutions of
Eq.~\eqref{eq:loc-eq} occurs, which is embodied in the
form~\eqref{eq:gfda} of the coefficients
$\overline{\alpha}^{\makebox{\tiny $(m)$}}_{t}$: {\it whichever} the
initial condition, due to the {\it Brownian motion} the solution
$\phi_{t}$ of Eq.~\eqref{eq:le} always picks a non-vanishing
component on the {\it ground state} (i.e.
$\overline{\alpha}^{\makebox{\tiny $(0)$}}_{t} \neq 0$ a.s., for any
$t > 0$); accordingly, the normalized solution $\psi_{t}$ converges
almost surely towards such a state. This is how the asymptotic
convergence of the general solution of Eq.~\eqref{eq:non-lin}
towards a Gaussian state occurs. The detailed analysis can be found
in~\cite{bd}.\vskip 0.3cm

\noindent \textsc{Discussion of the time regimes.} We conclude this
note by showing how one can read from~\eqref{eq:fgbxcv} the three
time regimes we mentioned at the beginning, which depict the main
features of the evolution of the solution $\psi_{t}$ of
Eq.~\eqref{eq:non-lin}. As discussed e.g. in~\cite{as4}, it is
physically convenient to take $\lambda$ proportional to the mass $m$
of the particle, according to the formula:
\begin{equation}
\lambda \; = \; \lambda_{0}\, \frac{m}{m_{0}},
\end{equation}
where $m_{0}$ is a reference mass and $\lambda_{0}$ is the universal
coupling constant which applies to all systems. Accordingly, $m$ is
the only free parameter of the model. The above assumption embodies
the fact that, for a composite system, the coupling constant
$\lambda{\text{\tiny cm}}$ associated to the motion of the center of
mass of the systems turns out to be the sum of the coupling
constants $\lambda_{i}$ associated to the motion of each
constituent. In models of wave function collapse---which are the
only ones we will refer to, in the following---in order for the
collapse mechanism to have the same strength of that of the original
GRW model~\cite{grw}, one has to set $\lambda_{0} \simeq 1.00 \times
10^{-2}$ m$^{-2}$ sec$^{-1}$ when $m_{0}$ is taken equal to the mass
of a nucleon ($\simeq 1.67 \times 10^{-27}$ Kg).

The first important point to stress is that the series
in~\eqref{eq:fgbxcv} is convergent only for $t
> \overline{t} := (4c +1)/\omega$; since $c$ is of order 1 and
$\omega \simeq 5.01 \times 10^{-5}$ sec$^{-1}$ (in a GRW-like
scenario), then $\overline{t} \simeq 2.00 \times 10^{4}$ sec. As we
shall soon see, this implies that the
representation~\eqref{eq:fgbxcv} is valid only for the classical and
diffusive regime, to which the long time behavior belongs. If one
wants to analyze the collapse regime and possibly the onset of
classical motion other representations of the solution must be used,
e.g. the one in terms of the Green's function. Given this, here we
shall use~\eqref{eq:fgbxcv} to describe all three time regimes,
keeping in mind that, for $t \leq \overline{t}$, the
representation~\eqref{eq:fgbxcv} makes sense only if a finite sum of
terms appears. This is nevertheless sufficient to describe,
qualitatively, the time evolution of the solution of
Eq.~\eqref{eq:non-lin}.

According to Eq.~\eqref{eq:jvcxz}, the $n$-th term of the
series~\eqref{eq:fgbxcv} defining $\phi_{t}$ decays exponentially in
a time equal to $1/n \omega$, which means that after a given time
$T$ all terms with $n > 1/ \omega T$ have decayed. This is the {\it
collapse regime}, previously introduced: any initially spread out
wave function collapses in space. However, according
to~\eqref{eq:dssd} the frequency $\omega$ depends on the ration
$\lambda/m$ and thus is independent of the mass of the particle:
this means that both for micro- and for macro-objects the
eigenstates decay in the same way; in particular, in a GRW-like
model, after a time $T \simeq 10^{-3}$ sec, which is about the
perception time of a human being, only the states with $n
> 2.00 \times 10^{7}$ have decayed. This is a very large number,
which seems not to justify the reason why, according the collapse
models, the wave function collapses the faster, the bigger the
system.

However, the parameter $z^2$ defined in~\eqref{eq:gfds}, which sets
the width of the eigenstates $u_{n}$ of the NSA harmonic oscillator
through Eq.~\eqref{eq:ev}, grows linearly with $m$. This implies
that the spread in position $\sigma_{q}^{n}$ of the state $u_{n}$
decreases with the mass according to the formula:
\begin{equation}
\left.\phantom{\sqrt{\frac{}{}}}\sigma_{q}^{n}\right|_{\text{\tiny
mass = $m$}} \; = \; \left.\sqrt{\frac{m_{0}}{m}}\,
\sigma_{q}^{n}\right|_{\text{\tiny mass = $m_{0}$}}.
\end{equation}
Accordingly, after a given time $T$, the states with $n < 1/ \omega
T$ which have survived the collapse, all have a spread, which is the
smaller, the bigger the system, and thus sum up to form a wave
function whose extension, for fixed $T$, decreases as $m$ increases.
This is how the amplification mechanism works, which says that the
{\it collapse regime} is shorter, the bigger the mass of the system.
This behavior can be seen also by inspecting directly
Eq.~\eqref{eq:non-lin}: if one performs the replacement $x
\rightarrow y = \sqrt{m/m_{0}}\, x$, then the dependence of the
equation on $m$ disappears. This means that the only effect of the
mass is to ``shrink'' the wave function in space, not to influence
directly its collapse in time. The fact that the collapse occurs in
good agreement with the Born probability rule is not so easy to see;
ref.~\cite{bd} contains a discussion of this point.

Let us now consider the case of a macroscopic particle and let us
suppose that the wave function has collapsed to a function which, on
macroscopic scales, is almost point-like. Then, within macroscopic
scales, the average value $\langle q \rangle_{t}$ and $\langle p
\rangle_{t}$ of the position and momentum operators, respectively,
are practically equal to $\overline{x}_{t}$ and $\hbar
\overline{k}_{t}$ defined in~\eqref{eq:prima2}
and~\eqref{eq:prima2b}, i.e. with the average position and momentum
of the Gaussian ground state. We can then analyze $\overline{x}_{t}$
and $\hbar \overline{k}_{t}$ in place of $\langle q \rangle_{t}$ and
$\langle p \rangle_{t}$; e.g. let us consider the time evolution of
$\overline{x}_{t}$.

As we see from Eq.~\eqref{eq:prima2}, the deterministic part of the
evolution corresponds to the solution of Newton's laws, while the
random part of the evolution is proportional to $m^{-1/2}$, thus it
is suppressed for large systems. E.g.: $\sqrt{\lambda} \hbar / m
\simeq 2.57 \times 10^{-19}$ m sec$^{-3/2}$ and $\sqrt{\hbar/m}
\simeq 3.24 \times 10^{-16}$ m sec$^{-1/2}$, for a 1-g object. The
figures show that for very long times the (almost point-like) wave
function moves practically along a straight line, like a classical
particle. This corresponds to the {\it classical regime}, where the
motion is essentially classical. This regime last the longer, the
bigger the system; for macroscopic systems, as the numerical example
shows, it is much longer than the time during which the system can
be kept isolated, so that its dynamics can be described by
Eq.~\eqref{eq:non-lin}.

The classical regime ends when the fluctuations become relevant on a
macroscopic scale, i.e.essentially when $W_{t} \sim \sqrt{m/\hbar}$
which is about $9.53 \times 10^{10}$ sec if we take Angstr\"om units
for length. After this time, the fluctuations start to blur the
deterministic motion: this is what we have called the {\it diffusive
regime}, which dominates the remaining evolution of the wave
function.

\vskip 0.3cm

\noindent \textsc{Acknowledgements.} We thank P. Pickl for very
useful discussions. The work was supported by the EU grant MEIF CT
2003--500543 and by DFG (Germany).

\end{document}